# Magnetic structure and multiferroicity of Sc-substituted hexagonal YbFeO$_3$

Y. S. Tang,[1] S. M. Wang,[1,] L. Lin,[1,*] V. Ovidiu Garlea,[2] Tao Zou,[3,†] S. H. Zheng,[1] H.-M. Zhang,[4] J. T. Zhou,[5] Z. L. Luo,[5] Z. B. Yan,[1] S. Dong,[4,‡] T. Charlton,[2] and J.-M. Liu[1]

[1]*Laboratory of Solid State Microstructures, Nanjing University, Nanjing 210093, China*

[2]*Neutron Science Division, Oak Ridge National Laboratory, Tennessee 37831, USA*

[3]*Collaborative Innovation Center of Light Manipulations and Applications, Shangdong Normal University, Jinan 250358, China*

[4]*School of Physics, Southeast University, Nanjing 211189, China*

[5]*National Synchrotron Radiation Laboratory, University of Science and Technology of China, Hefei 230026, China*

* llin@nju.edu.cn

† taozoucn@gmail.com

‡ sdong@seu.edu.cn

**Abstract**

Hexagonal rare-earth ferrite $R$FeO$_3$ family represents a unique class of multiferroics exhibiting weak ferromagnetism, and a strong coupling between magnetism and structural trimerization is predicted. However, the hexagonal structure for $R$FeO$_3$ remains metastable in conventional condition. We have succeeded in stabilizing the hexagonal structure of polycrystalline YbFeO$_3$ by partial Sc substitution of Yb. Using bulk magnetometry and neutron diffraction, we find that Yb$_{0.42}$Sc$_{0.58}$FeO$_3$ orders into a canted antiferromagnetic state with the Néel temperature $T_N \sim$ 165 K, below which the Fe$^{3+}$ moments form the triangular configuration in the *ab*-plane and their in-plane projections are parallel to the [100] axis, consistent with magnetic space group *P*6$_3$*c'm'*. It is determined that the spin-canting is aligned along the *c*-axis, giving rise to the weak ferromagnetism. Furthermore, the Fe$^{3+}$ moments reorient toward a new direction below reorientation temperature $T_R \sim$ 40 K, satisfying magnetic subgroup *P*6$_3$, while the Yb$^{3+}$ moments order independently and ferrimagnetically along the *c*-axis at the characteristic temperature $T_{Yb} \sim$ 15 K. Interestingly, reproducible modulation of electric polarization induced by magnetic field at low temperature is achieved, suggesting that the delicate structural distortion associated with two-up/one-down buckling of the Yb/Sc-planes and tilting of the FeO$_5$ bipyramids may mediate the coupling between ferroelectric and magnetic orders under magnetic field. The present work represents a substantial progress to search for high-temperature multiferroics in hexagonal ferrites and related materials.

# I. Introduction

Multiferroics, where the ferroelectric and magnetic orders are coupled with each other, have stimulated enormous attention not only because of a plenty of manifested exotic physical phenomena but also promising functionalities for novel applications [1-5]. However, most known multiferroics do not show ferroic ordering until extremely low temperature ($T$), which becomes a serious barrier for applications. As such, searching for high-$T$ multiferroics is one of the most vital tasks in the magnetoelectric (ME) community, but in a long period only rare systems can be qualified [6, 7].

Recently, it has been shown that hexagonal ferrite family $h$-$R$FeO$_3$ ($R$ = rare earth ions, Y), the so-called improper ferroelectrics, provides another approach to achieving high-$T$ multiferroicity [8-15], because of the stronger Fe$^{3+}$ spin exchanges in $R$FeO$_3$ compared to the Mn$^{3+}$ spin exchanges in $R$MnO$_3$. In this family, $h$-$R$FeO$_3$ is ferroelectric at room temperature, and ferroelectric (FE) polarization ($P$) is generally introduced by the structural distortion in association with transition from high-$T$ non-polar $P6_3/mmc$ phase to low-$T$ polar $P6_3cm$ phase, e.g. the FE transition occurs around 1000 K for $h$-LuFeO$_3$ film [11-12]. The cooperative tilting of FeO$_5$ bipyramids and buckling of $R$ layers lead to net electric polarization [13]. This finding seems to make the $h$-$R$FeO$_3$ family another set of multiferroics in addition to well-investigated $R$MnO$_3$ family. Nevertheless, different from $R$MnO$_3$ that favors orthorhombic structure as $R$ ion is relatively large and hexagonal structure as $R$ ion is small, free-standing $R$FeO$_3$ family all favors orthorhombic structure ($o$-$R$FeO$_3$), as shown in Fig.1(a).

Alternatively, next to the stable $o$-phase, hexagonal structure ($h$-phase) as the metastable one, as shown in Fig.1(b), can be obtained in thin film form if proper clamping from hexagonal substrate can be realized [11-15]. This allows an exploration of multiferroic behaviors of metastable $h$-$R$FeO$_3$. In fact, earlier work on $h$-LuFeO$_3$ films did receive attention, and the film sample first exhibited an antiferromagnetic (AFM) transition at $T = T_{N1} \sim 440$ K. Subsequently, a spin reorientation at $T = T_R = 130$ K appeared [11], and later neutron diffraction experiments suggested a ferromagnetically canted AFM ordering at the Néel point $T_N \sim 155$ K regardless of the film thickness [12]. This discrepancy might originate from the limited neutron diffraction data for determining the magnetic structure of film samples, appealing for more experiment on bulk systems. Fortunately, Masuno *et al.* found that the bulk $h$-phase could be stabilized upon Sc doping into Lu site for LuFeO$_3$, and consequently, $T_N$ was significantly enhanced, e.g. from 120 K in $h$-LuFeO$_3$ thin film [16] to ~172 K in $h$-Lu$_{0.5}$Sc$_{0.5}$FeO$_3$ bulk [17, 18], while the ferroelectricity remained robustly unaffected according to the first-principles calculations [18].

Most recently, a recorded $T_N \sim 185$ K was achieved by structural distortion strategy on some hexagonal ferrites [19]. In addition, large ME coupling or linear ME vortex structure in hexagonal ferrites was theoretically predicted, but not yet experimentally confirmed [20]. Even though the cloverleaf FE-vortex domains and large-loop weak ferromagnetic domains were observed in single crystal (Lu, Sc)FeO$_3$, no coupled ME effect between the FE domains and weak ferromagnetic domains has been demonstrated [21].

From the above statement, one can find that the $h$-$R$FeO$_3$ demonstrates more transitions than $h$-$R$MnO$_3$ in the magnetism and polarization due to the stronger Dzyaloshinskii-Moriya interaction between Fe$^{3+}$ ions under the formation of long-range magnetic ordering. It is worth noting that the $h$-$R$FeO$_3$ has a trimer distortion lattice at ferroelectric phase ($P6_3cm$) similar to the $h$-$R$MnO$_3$ features, while the particular Fe-Fe interaction is expected to instrinsically influence on trimerization by the spin-lattice coupling. Such as, earlier experiments on $h$-YbFeO$_3$ thin films reveal two-step FE transitions occurring at $T \sim 470$ K and $\sim 225$ K respectively, accompanied by pronounced magneto-dielectric effect near $T \sim 225$ K, while these properties are not available in $h$-$R$MnO$_3$. In addition, so far the 4$f$-3$d$ coupling in $h$-$R$FeO$_3$, if $R^{3+}$ is strongly magnetic, has not been paid sufficient attention for understanding the multiferroic behaviors, while this coupling is non-negligible or even very important for the multiferroicity in $R$MnO$_3$. Furthermore, this coupling can be remarkable far above the independent 4$f$ spin ordering point (usually < 10 K), making substantial contribution to the ME coupling at relatively high temperature [22-27]. More interestingly, the accordant steplike anomaly in magnetization and polarization at critical field in YbMnO$_3$ indicates strong association with a change in the nature of the field-induced spin reorientation below the Yb$^{3+}$ ordering [28]. All this inspires us to further study the magnetic structure and multiferroicity in $h$-$R$FeO$_3$ ($R^{3+}$ is magnetic ions, such as Yb$^{3+}$), which are expected as a nontrivial candidate for high-$T$ ME coupling and emergent physics beyond $h$-LuFeO$_3$ [13].

Earlier works on $h$-YbFeO$_3$ thin films did show quite a few inconsistencies regarding the FE transitions and ME coupling, and we highlight them here. First, are there indeed FE phase transitions occurring at $T \sim 470$ K and $\sim 225$ K? Second, is there any direct evidence with magnetic anomaly associated with these FE transitions? Third, how about the magnetism and ME effect in this system? While quite large magnetization at low $T$ was reported in Ref. [13], e.g., residual magnetization $M_r > 2$ $\mu_B$/f.u. at $T = 3$ K and $\sim 0.7$ $\mu_B$/f.u. at $T = 20$ K, and similar data were reported in Ref. [29], only very small value $M_r \sim 0.06$ $\mu_B$/f.u. at $T = 18$ K was probed in Ref. [30]. Indeed, it is challenging to detect the intrinsic magnetism of thin film samples for ferrites not only due to magnetic signals from $Fe_2O_3$-like impurities, if any, but also contributions from substrates, and thus an experimental study on bulk $h$-YbFeO$_3$ becomes urgently needed not only for clarifying these issues.

In this work, we start from synthesis of high-quality bulk $h$-YbFeO$_3$ polycrystalline samples by adopting substitution of Sc onto the Yb site. Using magnetic susceptibility and neutron powder diffraction measurements, we unveil convincingly three successive magnetic phase transitions at $T = T_N \sim 165$ K, $T = T_R \sim 40$ K, and $T = T_{Yb} \sim 15$ K, respectively. At $T_N \sim 165$ K, the system enters a canted AFM state with $Fe^{3+}$ moments lying in the $ab$-plane and a small canting moment along the $c$-axis giving rise to the weak ferromagnetism. Subsequently, the $Fe^{3+}$ spin projection in the basal plane reorients to a new direction at $T_R \sim 40$ K, followed by the independent ferrimagnetic $Yb^{3+}$ spin ordering along the $c$-axis below $T_{Yb} \sim 15$ K. The observed of electric polarization and dielectric anomaly at $T_N$ provides clear evidence for simultaneous magnetic and ferroelectric transitions, which is believed to attribute to the spin-lattice coupling [31]. In addition, remarkable ME effect in the low-$T$ range has been confirmed among the hexagonal ferrite family.

## II. Experimental details

The polycrystalline $h$-Yb$_{0.42}$Sc$_{0.58}$FeO$_3$ ($h$-YSFO) samples were synthesized using conventional solid-state reaction [9]. Stoichiometric mixtures of high purity $Yb_2O_3$, $Sc_2O_3$, and $Fe_2O_3$ were thoroughly ground, and sintered at 1200 °C for 24 hours. Then the powder was pelletized and sintered at 1400 °C for 24 hours with intermittent grindings.

The crystallinity of the as-prepared samples was checked using the X-ray diffraction (XRD, D8 Advanced, Bruker) in the $\theta$-$2\theta$ mode with Cu K$_\alpha$ source ($\lambda$ = 1.5406 Å) at room temperature. To monitor the $T$-dependent structural evolution, *in situ* high-resolution X-ray diffraction was performed at the 14B beamline of Shanghai synchrotron radiation facility (SSRF) with a Linkam cryo-stage. Neutron powder diffraction (NPD) data were collected at HB-2A powder diffractometer at Oak Ridge National Laboratory, USA. The wavelength of $\lambda$ = 1.54 Å was used at room temperature for the refinement of nuclear structure, while $\lambda$ = 2.41 Å was used at low temperatures for the refinement of both nuclear and magnetic orderings. The refinement was carried out using the Rietveld analysis program package FULLPROF [32]. The symmetry allowed magnetic models were analyzed using the software package SARAh [33]. To check the samples' stoichiometry, the electron dispersion spectroscopy (EDS) (Quanta 200, FEI) was used to analyze the chemical composition, noting that the $Fe^{3+}$ valence in $(Lu_{1-x}Sc_x)FeO_3$ is highly preferred [18].

To measure dielectric and ferroelectric properties, a sandwich-type capacitor was made by depositing Au on the top/bottom surfaces of the disk-like sample with 3.0 mm in diameter and ~ 0.2 mm in thickness. The $T$-dependence of dielectric constant ($\varepsilon$) was measured using the HP4294A impedance analyzer (Agilent Technologies, Inc.) integrated with Physical Property Measurement System (PPMS) (Quantum Design, Inc.). The specific heat ($C_P$) was measured using the PPMS in the standard procedure. The magnetic susceptibility ($\chi$) as a function of $T$ under the zero-field cooling (ZFC) and field cooling (FC) modes was measured by the Superconducting Quantum Interference Device Magnetometer (SQUID) (Quantum Design, Inc.), using the measuring/FC field of 1.0 kOe. The isothermal magnetic hysteresis (magnetization $M$ vs magnetic field $H$) was also probed in the standard mode.

The conventional pyroelectric current method was used to probe the ferroelectric polarization [34, 35]. In detail, the sample was cooled down from 180 K to 10 K under a poling electric field ($E_p$ = 10 kV/cm). Subsequently, the electric field was removed, followed by sufficiently long short-circuit to reduce the background of electrical current with minimum level (i.e., < 0.2 pA). The extrinsic contributions, e.g. trapped charges during the electric field poling, or charges contributed from thermal activation during warming process were carefully examined. The $T$ dependent pyroelectric current $I_p$ was collected with the heating rate of 4 K/min. In addition, the magnetic field-driven polarization ($\delta P$) was obtained by integrating the magnetoelectric current $I_M(H)$, which was measured using the Keithley 6514 electrometer upon increasing $H$ from $H$ = 0 to $H$ = ± 7 T at ramping magnetic field rate 100 Oe/s after the poling and the short-circuit process.

### III. Results and discussion
*A. Crystal structure*

For the sample synthesis, a series of YSFO ($0.3 \leq x \leq 1.0$) samples were prepared in order to obtain pure hexagonal phase. At low $x$ ($x \leq 0.6$), the samples can be clearly indexed as the combinations of $o$-$Yb_{1-x}Sc_xFeO_3$, $h$-$Yb_{1-x}Sc_xFeO_3$, and very tiny $(Yb_{1-x}Sc_x)_3Fe_5O_{12}$ [36]. A further increasing of $x$ makes the intensity of Bragg peaks of $o$-$Yb_{1-x}Sc_xFeO_3$ phase to gradually drop down, and almost disappear in sample $x$ = 0.6 where the $h$-$YbFeO_3$ phase is dominated, accompanied with tiny amount of bixbyite phase $ScFeO_3$. For higher $x$ (> 0.7), the bixbyite phase (see Fig. 1(c)) gradually increases.

It is more challenging to obtain pure $h$-YSFO phase in comparison with previously synthesized $h$-Lu$_{1-x}$Sc$_x$FeO$_3$ [17, 18]. The larger Yb$^{3+}$ ions are disadvantageous to stabilize the hexagonal structure than Lu$^{3+}$ and the ionic size mismatch between Yb$^{3+}$ and Sc$^{3+}$ is more obvious. These are non-favored for obtaining pure $h$-phase and the ionic occupation disordering must be more remarkable. According to the XRD pattern shown in Fig. 1(d), sample $x = 0.58$ has the purest $h$-phase, with only very tiny bixbyite impurity, as indicated in the inset. The tiny ScFeO$_3$ as impurity can be safely neglected in our discussion since it is paramagnetic above $T \sim 20$ K [37] and non-ferroelectric due to the cubic space group $Ia$-3 [38]. The high-quality data refinement shows that the lattice structure of sample $x = 0.58$ is in good agreement with hexagonal symmetry and the lattice constants are $a = b = 5.8523$Å and $c = 11.7008$ Å.

The chemical compositions were checked using the EDS measurement (Fig. 1(e)) and the data show that the atomic ratio Yb:Sc:Fe is 1.0:1.182:2.270. While the sum of Yb and Sc is lower (by 4%) than the Fe content, it is enough safe to conclude that the samples are stoichiometric considering the uncertainties of EDS technique.

*B. Magnetic phase transitions*

We first check the magnetic phase transitions. The *dc* magnetic susceptibility $\chi(T)$ for sample $x = 0.58$ in the $T$-range from 2 K and 300 K under a measuring field $H \sim 1$ kOe is plotted in Fig. 2(a). It is seen that the sample first undergoes a weak anomaly around $T = T_C \sim 225$ K, suggesting some magnetic ordering event. The $\chi(T)$ curves under the ZFC and FC modes have a weaker anomaly at $T_C$, while the splitting emerges below $T_N$. Upon further cooling, a cusp feature emerging at $T = T_N \sim 165$ K was clearly identified, indicating another magnetic phase transition. At $T = T_R \sim 40$ K, a small bump shows up in the $\chi(T)$ curve, suggesting third magnetic ordering event. For checking the nature of these anomalies, the specific heat divided by temperature $C_P/T$ was measured as shown in Fig. 2(b). Apparently, the AFM transition at $T_N$ is confirmed by an anomaly, while no identifiable anomaly at $T_C$ can be detected, raising question on the long-range ordering nature of the event at $T_C$.

One may also plot the $(d\chi/dT) \sim T$ curve in Fig. 2(a), and identifies the three anomalies at $T_C$, $T_N$, and $T_R$. However, the feature at $T_C$ is very dim, suggesting no long-range magnetic ordering. Considering the fact of no anomaly at $T_C$ in the $C_P(T)$ curve and splitting of the $\chi(T)$ curves under the ZFC and FC modes below $T_C$, it is argued that most likely there appears ferromagnetic (FM)-like clustering just below $T_C$ and these clusters don't merge together until $T \to T_N$, below which a long-range AFM ordering develops. Due to this long-range AFM ordering, one sees a significant splitting between the ZFC and FC modes below $T_N$, which reveals a weaker ferromagnetic ordering.

In proceeding, we can evaluate the magnetic interactions by fitting the $\chi(T)$ data under the ZFC mode, as shown in Fig. 2(c) where the $\chi^{-1}(T)$ curve is plotted as well. The Curie-Weiss fitting of the paramagnetic data between 350 K ~ 400 K gives a negative Curie-Weiss temperature $\theta_{cw}$ ~ -185 K and an effective moment $\mu_{eff}$ ~ 4.8 $\mu_B$/f.u, indicative of strong AFM interaction. Furthermore, the magnetic transition at $T_R$ ~ 40 K identified here was also reported for the $h$-Lu$_{1-x}$Sc$_x$FeO$_3$ single crystal [17]. It is noted that all the $h$-$R$MnO$_3$ and those $h$-$R$FeO$_3$ containing Sc samples exhibit spin reorientation at low temperature, e.g., $T$ = 37 K for $h$-HoMnO$_3$ [39], 43.5 K for $h$-YbMnO$_3$ [40], 40 K for $h$-LuMnO$_3$ [41], 35 K for $h$-YMn$_{0.9}$Fe$_{0.1}$O$_3$ [42], and 45 K for $h$-Lu$_{0.5}$Sc$_{0.5}$FeO$_3$ [17]. The earlier neutron scattering provided clear evidence for the Mn/Fe ionic shift from the 1/3 position [17, 43]. Such shift does not suppress the rotational invariance in the Mn plane but clearly lifts the inter-plane frustration, thus leading to remarkable variations of the in-plane and out-of-plane exchange and magnetocrystalline anisotropy. The spin reorientation is believed to be related with these variations at $T_{SR}$. In contrast, the reported AFM ordering point $T_N$ of $h$-YbFeO$_3$ is ~ 120 K, and a spontaneous magnetization reversal at $T$ ~ 83 K was claimed due to the competition between two magnetocrystalline anisotropy terms associated with Fe$^{3+}$ and Yb$^{3+}$ moments [13].

The macroscopic magnetism can be further characterized by measuring the $M$-$H$ hysteresis, as shown Fig. 3(a) for several selected temperatures. The FM-like hysteresis is identified below $T_N$, while the magnetic clusters can give rise to the weak hysteresis loops above $T_N$ until $T$ ~ $T_C$. To better quantify the magnitude of its effects, the specific $M$-$H$ curves near $T_C$ with magnetic field between -0.2 T to 0.2 T are plotted in Fig. 3(b). We can see significant hysteresis loops below $T_C$ point, and a weak residual moment ($M_r$) ~ 0.002 $\mu_B$/f.u. at $T$ ~ 200 K is clearly demonstrated. The evaluated residual moment as a function of $T$ are summarized in Fig. 3(c). Here it is noted that there is minor garnet impurity phase at low doping content. The garnet phase is ferrimagnetic with about 1.0 $\mu_B$/Fe [44]. Therefore, even 1% of this phase can add 0.01 $\mu_B$/Fe to the total magnetization. Although we did not observe any magnetic impurity within resolution of XRD instrument, the possibility of the garnet impurity phase as the source of magnetic clusters can not be excluded. Previous first-principles calculations have shown that there is a small spin canting out of the $xy$ plane that leads to a net magnetization $M_z$=0.02 $\mu_B$/Fe for $h$-LuFeO$_3$ along the $z$-axis, displaying weak ferromagnetism [20]. Our $M$-$H$ data supports the theoretical prediction. The measured moment ($M_r$) varies from 0.01 $\mu_B$/f.u. at $T$=100 K to 0.05 $\mu_B$/f.u. at $T_R$, below which the net magnetization increases dramatically due to the introduction of $Yb^{3+}$ ordering. Therefore, these behaviors qualitatively agree with theoretical prediction of weak but intrinsic ferromagnetism generated from the canted $Fe^{3+}$ spins that order antiferromagnetically. It should be mentioned here that the weak FM hysteresis in *h*-YbFeO$_3$ thin films at low *T* was already reported [13], and the measured residual moment is as large as ~ 2.0 $\mu_B$/f.u.. Such a large moment should not be intrinsically associated with the spin-canting, noting that the AFM order is highly favored by the very negative $\theta_{cw}$ ~ -185 K. In the present work, our samples show much smaller residual moment which is ~ 0.18 $\mu_B$/f.u. at $T$ ~ 5 K and ~ 0.09 $\mu_B$/f.u. at $T$ ~ 20 K, comparable with the reported value of ~ 0.06 $\mu_B$/f.u. at $T$ ~ 18 K [30]. It is thus well confirmed that the system exhibits the AFM order below $T_N$ with canted spin moment, and it is also proposed that the system contains some FM-like clusters in $T_N < T < T_C$. These will receive further checking by the neutron scattering investigation to be shown below.

*C. Neutron diffraction for magnetic structure*

For determining the magnetic structure, we performed the NPD measurements on the samples at several selected temperatures. The neutron wavelength for probing at $T$ = 300 K is $\lambda$ = 1.54 Å, and that at $T$ = 200 K, 125 K, 25 K, and 3 K is $\lambda$ = 2.41 Å, noting that the temperature choice was made referring to the values of $T_C$, $T_N$, $T_R$ etc.

As shown in Fig. 4(a), all the Bragg reflections at $T$ = 300 K can be well indexed with the hexagonal space group $P6_3cm$. The refined lattice constants are $a = b$ = 5.86054(8) Å, $c$ = 11.70606(22) Å, respectively, consistent with the XRD data shown in Fig. 1(d). The most important crystallographic information extracted from the refinement and the related discrepancy factors are shown in Table. I. The cubic bixbyite-type ScFeO$_3$ (space group $Ia$-3) was added as the second phase for the refinement and it shows that this tiny phase occupies ~ 3.8(3) wt% in amount at most.

Now we can discuss the magnetic structures associated with the various phases below $T_C$, $T_N$, and $T_R$, respectively, by combining the neutron scattering data and magnetometry results. First, no magnetic peaks were detected in addition to the nuclear ones at $T$ ~ 200 K and above. We note that a ferromagnetic component of less than 0.1 $\mu_B$ is nondetectable using unpolarized neutrons, and therefore our neutron measurements cannot probe the moment increase occuring around $T_C$ ~ 225 K, where the remnant moment is smaller than 0.01 $\mu_B$. Second, we collected the NPD data at $T$ = 125 K, 25 K, and 3 K respectively. The data and their refinement results at the three temperatures are plotted in Fig. 4(b), (c), and (d). Several issues can be clarified here before we discuss the magnetic structures at various temperatures:

(1) No satellite reflection was observed, revealing that the magnetic unit cell is identical to the nuclear (chemical) one. The magnetic propagation vector $k = (0, 0, 0)$ was confirmed using the $k$-search function of the FullPROF package, consistent with other hexagonal manganites and ferrites [12].

(2) Symmetry allowed magnetic models were analyzed based on the irreducible representation analysis using the SARAh software [33]. Four corresponding magnetic structure models are derived from the four irreducible representations $\Gamma_1$, $\Gamma_2$, $\Gamma_3$ and $\Gamma_4$ of symmetry $P6_3cm$ and $k = (0\ 0\ 0)$.

(3) It is noted that the only irreducible representation which gives rise to FM component along the $c$-axis is $\Gamma_2$, equivalent to the magnetic space group $P6_3c'm'$.

(4) Furthermore and even more important, we collected the intensity evolution data of the magnetic peaks in the low-$Q$ region as a function of $T$ and the $T$-dependent contour map is plotted in Fig. 4(e). To clearly show the transition temperatures, the peak intensities of (101), (100) and (102) are also plotted as function of $T$ respectively, as shown in Fig. 4(f). These data present the basis for us to evaluate the magnetic structures in various $T$-ranges.

We first address the magnetic structure between $T_R < T < T_N$. As shown in Fig. 4(e) and (f), it is clearly identified that (101) peak which is structurally forbidden by the $P6_3cm$ space group symmetry for nuclear diffraction emerges at $T_N = 165$ K, indicating its entire magnetic origin. In addition, the $M$-$H$ hysteresis does show minor FM-like loop below $T_N$, due to the spin canting along the $c$-axis, as evidenced in Fig. 3(a), and this effect is similar to the case of $h$-LuFeO$_3$ [12]. Since only $\Gamma_2$ ($P6_3c'm'$) allows the existence of a ferromagnetic component along $c$-axis, it has to be included to refine the magnetic structure. On the other hand, the irreducible representations $\Gamma_1$ and $\Gamma_3$ can be safely ruled out since they must contribute to the (100) peak which however remains absent until $T_R$. The best refinement can be reached with a nuclear model in combination with a magnetic structure characterized by $\Gamma_2$, as shown in the Fig. 4(b). The evaluated magnetic structure for $Fe^{3+}$ moment is schematically drawn in Fig. 5(a). The refined magnetic moments are $m_a = m_b = 2.511\ \mu_B$, $m_c = 0.783\ \mu_B$, $m_{tot} = 2.63\ \mu_B$, and the magnetic R-factor is 15.9.

Towards $T < T_R \sim 40$ K, another magnetic peak (100) appears by remarkable increase in intensity with decreasing $T$, as shown in Fig. 4(e) and (f). Meanwhile, the intensity of (101) peak experiences an abrupt drop below $T_R$, accompanied by an obvious intensity increase of (102). The appearance of (100) and lowering of (101) intensity can be explained either by Fe rotation in the plane or from the ordering of Yb moment. The model based on $\Gamma_2$ with ordered Yb moments can be used to fit the data, but larger $Fe^{3+}$ moments than 3 K are acquired. This model is thus discarded. It is known that the irreducible representations $\Gamma_1$ and $\Gamma_3$ usually contribute significantly to the (100) peak. An emergence of the peak (100) at $T_R$ simply implies a reorientation of the $Fe^{3+}$ spins toward a new direction in the basal plane characterized by either $\Gamma_1$ or $\Gamma_3$. The reorientation is also captured by the anomaly around 40 K in the *M-T* curve. Considering the observed hysteresis loop below $T_R$, $\Gamma_2$ has to be included for the refinement. A magnetic model composed of basis vectors from two representations, $\Gamma_1$ and $\Gamma_2$, has given us the best refinement result, as shown in Fig. 4(c). This model is equivalent to the magnetic space group $P6_3$ (#173.129), and the magnetic structure for $Fe^{3+}$ moment is schematically drawn in Fig. 5(b). The refined moments are $m_a = 3.532\ \mu_B$, $m_b = 2.22\ \mu_B$, $m_c = 0.796\ \mu_B$, $m_{tot} = 3.193\ \mu_B$, and the magnetic R-factor is 16.1.

Upon further cooling to 15 K, the integrated intensity of (101) and (102) shows an obvious increase, accompanied with a minor decrease of (100) peak, as shown in Fig. 4 (f). This feature is very similar to isostructural $HoMnO_3$, where $Ho^{3+}$ magnetic moments order below $T = 25.4$ K [45]. Since rare earth $Yb^{3+}$ ions are also magnetic, it is reasonable to take the ordering of $Yb^{3+}$ moments into consideration for the refinement at 3 K. The refinement using the same magnetic symmetry at 25 K with ordered $Yb^{3+}$ moments give us the best refinement results. At $T \sim 3$ K, the $Fe^{3+}$ moment components are $m_a = 3.344\ \mu_B$, $m_b = 1.4\ \mu_B$, $m_c = 1.105\ \mu_B$, and the total moment is $m_{tot} = 3.111\ \mu_B$, while for $Yb^{3+}$ ions on site (2$a$), $m_c = 1.802\ \mu_B$, $Yb^{3+}$ ions on site (4$b$), $m_c = -1.511\ \mu_B$. The magnetic R-factor is 18.8.

While for (102), the intensity remarkably increases at $T_N$ and below due to the magnetic contribution. This is consistent with the specific heat peak around 165 K, as shown in Fig. 2(b), suggesting the onset of intrinsic magnetic phase transition at this temperature. In addition, we find that the intensity curve of (102) combines the features of the (101) and (100) reflections, indicating the involvement of magnetic contribution from both $Fe^{3+}$ and $Yb^{3+}$, similar to the case of $YbMnO_3$ [28]. In $R$FeO$_3$ and $R$MnO$_3$ systems, it is well known that the $R^{3+}$-$Fe^{3+}$/$R^{3+}$-$Mn^{3+}$ coupling is strong and the $Fe^{3+}$/$Mn^{3+}$ spin ordering may induce the $R^{3+}$ spin ordering too as the concurrent sequence. In this scenario, one may intuitively argue that the magnetic order is driven by the competing exchanges in this system, a very common phenomenon for magnetic oxides.

Besides the low-$T$ data, the NPD patterns were collected up to 550 K. The refinements at various temperatures reveal that the polar hexagonal structure remains unchanged in the entire measured temperature range, as indicated by the variation of lattice constants as functions of $T$ as shown in Fig. 4(g). It thus confirms that ferroelectric phase persists above $T \sim 550$ K. Previous density functional theory (DFT) calculations has explained the stability of the hexagonal phase in $h$-LuFeO$_3$ upon Sc substitution, while the multiferroic properties, including the noncollinear magnetic order and ferroelectricity remains robustly unaffected [18]. Under the ferroelectric state, $h$-Yb$_{0.42}$Sc$_{0.58}$FeO$_3$ belongs to non-centrosymmetric $P6_3cm$ space group at room temperature, similar to the isomorphic $h$-$R$MnO$_3$. Here, it is worth noting that Disseler *et al* have summarized the Néel transition temperatures as a function of the $c/a$ ratio lattice parameters for hexagonal $R$MO$_3$ ($R$=Lu, Dy, Sc, Y and $M$=Mn or Fe) [17]. They find a linear trend that is independent of both the $R$ species as well as the transition metal (Fe or Mn). Instead, $T_N$ is highly dependent of the ratio of $c$ to $a$. The $c/a$ ratio obtained from our neutron diffraction is 1.997 (300 K) and 1.998 (5 K), compared with that in $h$-Lu$_{0.5}$Sc$_{0.5}$FeO$_3$ (1.998 at 300 K) [9]. Therefore, the similar $c/a$ ratio between $h$-Lu$_{0.5}$Sc$_{0.5}$FeO$_3$ and $h$-Yb$_{0.42}$Sc$_{0.58}$FeO$_3$ confirms that the AFM transition at $T_N \sim 165$ K is intrinsic [36].

To this end, the magnetic space groups used to refine the magnetic structures are summarized in Table II, and the refined magnetic structures are plotted in Fig. 5.

### D. Ferroelectricity and Magnetoelectric Coupling

ME coupling effect in the multiferroics is typically manifested as control ferroelectric polarization ($P$) or magnetization ($M$) by applied magnetic field ($H$) or electric field ($E$). In some type-II multiferroics, e.g., TbMnO$_3$ and DyMnO$_3$, the magnetodielectric effect is also accompanied by a magnetoelastically induced lattice modulation, which results in the emergence of ferroelectric polarization. In fact, while most earlier works on hexagonal ferrites focused on nonmagnetic Lu$^{3+}$/Y$^{3+}$ based systems, indeed no direct observation of ME effect, e.g., magnetic field control of polarization, have yet been confirmed. One may recall previous theoretical work on the trimer structural distortion that induces not only a spontaneous polarization but also bulk magnetization and linear ME effect [20]. As $h$-Yb$_{0.42}$Sc$_{0.58}$FeO$_3$ is already ferroelectric at room temperature, we are very interested in the low-$T$ antiferromagnetic phase. Based on the magnetic structures evaluated above, one can now discuss the possible ferroelectricity associated with the magnetic structure, noting that both the magnetic space groups $P6_3c'm'$ and $P6_3$ are polar and allow the ME effect. In our $h$-YSFO system, the strong

Yb magnetism and $Fe^{3+}$-$Fe^{3+}$/$Fe^{3+}$-$Yb^{3+}$ coupling may make some difference, allowing our attention on the ferroelectric and dielectric properties of our sample subsequently.

It is verified that the polar hexagonal lattice structure remains unchanged in the whole covered $T$-range up to 550 K. Thus, the sample is already in the ferroelectric state at room temperature. Here, we focus our attention on the potential correlation between magnetism and ferroelectricty. Using the high-precision pyroelectric current method, we determined the change of polarization $\Delta P$ (not the total polarization) as a function of $T$ below room temperature. To proceed, a poling electric field ($E_p$ = 10 kV/cm) was applied from $T \sim$ 180 K to $T \sim$ 10 K, and the corresponding pyroelectric current ($I_p$) and integrated polarization $\Delta P$ are presented in Fig. 6(a). It is found that pyroelectric current $I_p$ begins to emerge just around $T_C$ (with a small shoulder). Upon cooling from $T_N \sim$ 165 K, a broad and large peak appears with the peak-location at $T \sim$ 120 K. This broad peak does not end until $T = T_R \sim$ 40 K, and thus a polarization change $\Delta P$ as large as $\sim$ 2.0 μC/cm$^2$ is obtained. Similar to that in $R$MnO$_3$, the ferroelectric-transition temperature ($T_{FE}$) is about 570-990 K, which is a very broad polarization region, while it has been reported that the structural-transition occurs at $T_s \sim$1350 K [46].

One may argue that the $I_p$ feature should be magnetically induced, considering the broad $I_p(T)$ peak covering the $T$-range from $T_N$ to $T_R$. In particular, the dielectric constant as a function of $T$, as plotted in Fig. 6(b) for zero magnetic field and $H$ = 5 T, evidences a clear bump around $T_N$, indicating the ME effect via the spin-lattice coupling. In addition, the dielectric loss is as small as 10$^{-3}$ on the order of magnitude for the whole $T$ range, indicating that the sample is highly insulating. It is thus speculated that the broad $I_p$ peak below $T_N$ originates from the AFM $Fe^{3+}$ spin ordering via the spin-lattice coupling. Let us clarify that the magnetism undergoes some change in this interval. A weak ferromagnetism with a net $m_c$=0.783 $\mu_B$ at $T$=125 K along the $c$ axis is formed. As $T$ reducing, the moment has an obvious change in plane from $m_a$ = 2.511 $\mu_B$ to $m_a$ = 3.532 $\mu_B$ at $T$=125 K and 25 K, respectively, and the total moment increases significantly, while the $m_c$ nearly unchanged. It is reasonable to understand that the polarization appearing in the $T_N$ is due to the stronger interaction of $Fe^{3+}$ result in lattice distortion, similar to the trimer distortion from the paraelectric to ferroelectric phase. The moment of in-plane changes in succession, reveals that the spin-lattice coupling may continue to the $T_R$. This proposed scenario also explains why the $I_p$ peak in Fig. 6(a) and thus the $\Delta P$ transition region are so broad.

To shed more light on the ME effect, we performed the iso-thermal ME measurements by probing the magnetoelectric current $I_M$ in response to the magnetic field and magnetic field-driven ferroelectric polarization ($\delta P$) at $T$ = 2 K, as shown in Fig. 7(a) where the arrows indicate the direction of current change. The measured $I_M \sim H$ butterfly loop represents the typical ME response, demonstrating the detectable ME effect. The coercive field corresponding to the ME signal reversal is roughly $H_c \sim 0.5$ T. Unfortunately, the measured ME current $I_M$ seems to be small and the evaluated polarization response $\delta P$ is on the order of magnitude of $\mu C/m^2$, much smaller than the polarization itself ($\Delta P \sim 2~\mu C/cm^2$) shown in Fig. 6(a). Such a small $\delta P$ is surely due to the robustness of the $Fe^{3+}$ AFM order against magnetic field up to 7.0 T. It is also reasonable since the spin-lattice mechanism generated ferroelectric polarization always shows weak ME response, like the cases of multiferroic $R$MnO$_3$.

This ME effect is significant between ±4 T. In Figs. 7(b) and 7(c), the ME effect repeats well with $H$ oscillating. The maximum of $\delta P$ is about -0.9 $\mu C/m^2$, indicating the coupling is weak. Considering the polycrystalline nature of the sample, the ME coupling is often week and might be obscure if the signal is intrinsically low, or if the magnetic energy gain is not sufficient to overcome the energy barrier between multiple grains. Nevertheless, it should be mentioned that a varied $\delta P$ of 8 $\mu C/m^2$ was observed in single crystal $h$-YbMnO$_3$ [28], thus the value in our case is considered as intrinsic signal. Hence, high-quality of single crystal hexagonal rare-earth ferrite is highly recommended to quantify the ME coupling phenomena by applying magnetic field along the different crystallographic directions, and unveil the underlying physical origin that is highly correlated to the $Fe^{3+}$ ordering, and $Fe^{3+}$-$R^{3+}$ interactions.

*E. Discussion*

We note that the ferrimagnetic phase is characterized by a linear ME effect as shown in Fig. 7(c). According to the NPD analysis, the magnetic point group is $P6_3c'm'$ below 165 K, and then evolves into $P6_3$ at 40 K. Both $P6_3c'm'$ and $P6_3$ are polar, and allow $c$-axis spontaneous polarization, consistent with previous report [13]. Furthermore, both $P6_3c'm'$ and $P6_3$ allow linear magnetoelectric effect from the space group theory. Nevertheless, our ME coupling effect exhibited in polycrystalline sample still shed more light on the hexagonal rare-earth ferrite $R$FeO$_3$, thus high quality of single crystal is highly required.

In addition, the high-resolution $T$-dependent synchrotron XRD was conducted to monitor the expected structure transition in our Yb$_{0.42}$Sc$_{0.58}$FeO$_3$ around the ferromagnetic transition point $T_C$. As shown in the XRD patterns in Figs. 8(a) and 8(b), there is no phenomenon such as peak splitting, appearing or vanishing, indicating none obvious phase transition occur in the $T$ range of 200~250 K, consistent with the neutron data. Even though, obvious changes of peak amplitudes of (008) and (222) planes are evidenced around $T_C$, as shown in Figs. 8(c) and 8(d). The $T$-dependent peak positions of selected crystal planes are presented in Figs. 8(e) and 8(f). Besides the overall decreasing trend caused by thermal expansion, clear turning points could be noticed in the curves belonging to (115), (008), and (222) of $Yb_{0.42}Sc_{0.58}FeO_3$. These result shows that a mild structural transition does occur around $T_C$, which is in good agreement with the electrical properties of $Yb_{0.42}Sc_{0.58}FeO_3$.

## IV. Conclusion

In summary, we have achieved good description of the magnetic properties and magnetoelectric coupling in bulk $h$-$Yb_{1-x}Sc_xFeO_3$ though neutron diffraction and electrical measurements. We find that the canted AFM state with $Fe^{3+}$ moments lying in the $ab$-plane emerges at $T_N$ = 165 K. Upon cooling to $T_R$ = 40 K, the $Fe^{3+}$ moments reorient toward a new direction, while the $Yb^{3+}$ moments order ferrimagnetically along the $c$-axis below the characteristic temperature $T_{Yb}$ ~ 15 K. Direct experimental observation of ME coupling was first verified in this hexagonal ferrite, which provide a promising candidate for hunting for multiferroics in other hexagonal $R$FeO$_3$ system and related materials. Moreover, high quality of single crystals is highly required to distinguish the magnetic configuration of two sites of the $Yb^{3+}$ ions, in particular when does the two $Yb^{3+}$ ions order independently, and which is expected to play an important role in the ME coupling that is associated with any possible field-induced metamagnetic transitions.

**Acknowledgment**

This work was supported by the State key Research Program of China (Program No. 2016YFA0300101, and 2016YFA0300102), the National Natural Science Foundation of China (Grant Nos. 11874031, 11834002, 11774106, 51721001, and 11974167). Z. L. Luo thanks the staff at beamline BL14B of SSRF for their support. The research at Oak Ridge National Laboratory's High Flux Isotope Reactor was sponsored by the Scientific User Facilities Division, Office of Basic Energy Sciences, US Department of Energy.

TABLE I. Refined structure parameters from powder neutron diffraction data measured at 300 K.

| Atom (Wyck.) | x | y | z | B |
|---|---|---|---|---|
| Yb1 (2a) | 0.0000 | 0.0000 | 0.26847(63) | 0.552(111) |
| Sc1 (2a) | 0.0000 | 0.0000 | 0.26847(63) | 0.552(111) |
| Yb2 (4b) | 0.33333 | 0.66667 | 0.23632(54) | 0.501(61) |
| Sc2 (4b) | 0.33333 | 0.66667 | 0.23632(54) | 0.501(61) |
| Fe1 (6c) | 0.33333 | 0 | 0.00000 | 0.025(28) |
| O1 (6c) | 0.33061 | 0 | 0.17040 | 0.522(76) |
| O2 (6c) | 0.63231 | 0 | 0.33802 | 0.139(12) |
| O3 (2a) | 0.0000 | 0.0000 | 0.47629 | 0.201(20) |
| O4 (4b) | 0.33333 | 0.6667 | 0.01236 | 0.678(68) |

**Space Group:** $P6_3cm$, $a = b = 5.86054(8)$ Å, $c = 11.70606(22)$ Å, $\text{Chi}^2 = 10.6$, $R_{\text{Bragg}} = 3.61$, $R_p = 3.73$, $R_{wp} = 4.84$

TABLE II. Magnetic space group and models used to refine the magnetic structures at 125 K, 25 K and 3 K. The magnetic subgroup is $P6_3c'm'$ #(185.201) at 125 K, and $P6_3$(#173.129) at 25 K and 3 K.

| $T = 125$ K | $T = 25$ K, $T = 3$ K |
|---|---|
| $(x,0,z \mid m_x,0,m_z)$ | $(x,0,z \mid m_x, m_y, m_z)$ |
| $(0,x,z \mid 0,m_x,m_z)$ | $(0,x,z \mid -m_y, m_x-m_y, m_z)$ |
| $(-x,-x,z \mid -m_x,-m_x,m_z)$ | $(-x,-x,z \mid -m_x+m_y,-m_x,m_z)$ |
| $(-x,0,z+1/2 \mid -m_x,0,m_z)$ | $(-x,0,z+1/2 \mid -m_x,-m_y,m_z)$ |
| $(0,-x,z+1/2 \mid 0,-m_x,m_z)$ | $(0,-x,z+1/2 \mid m_y,-m_x+m_y,m_z)$ |
| $(x,x,z+1/2 \mid m_x,m_x,m_z)$ | $(x,x,z+1/2 \mid m_x-m_y,m_x,m_z)$ |

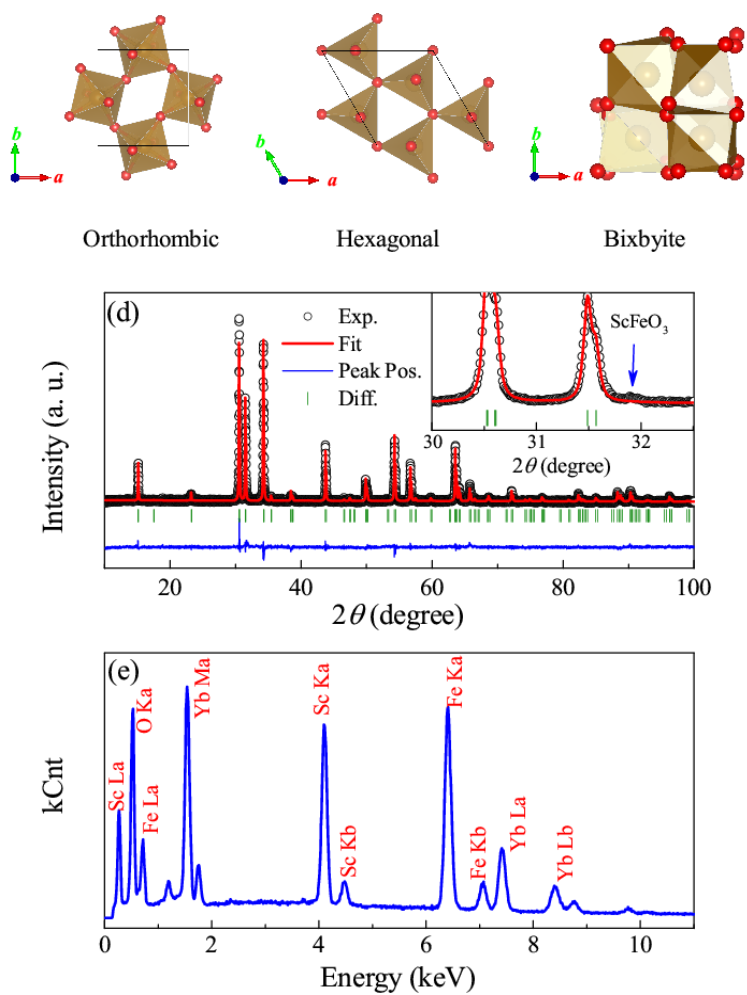

Fig. 1. The crystal structure of (a) orthorhombic, (b) hexagonal, and (c) bixbyite $R$FeO$_3$ ferrites. (d) Rietveld profile fitting result for the XRD pattern of Yb$_{0.42}$Sc$_{0.58}$FeO$_3$. The tiny impurity of ScFeO$_3$ is indicated in the inset. (e) The cation/anion stoichiometry indicated by EDS.

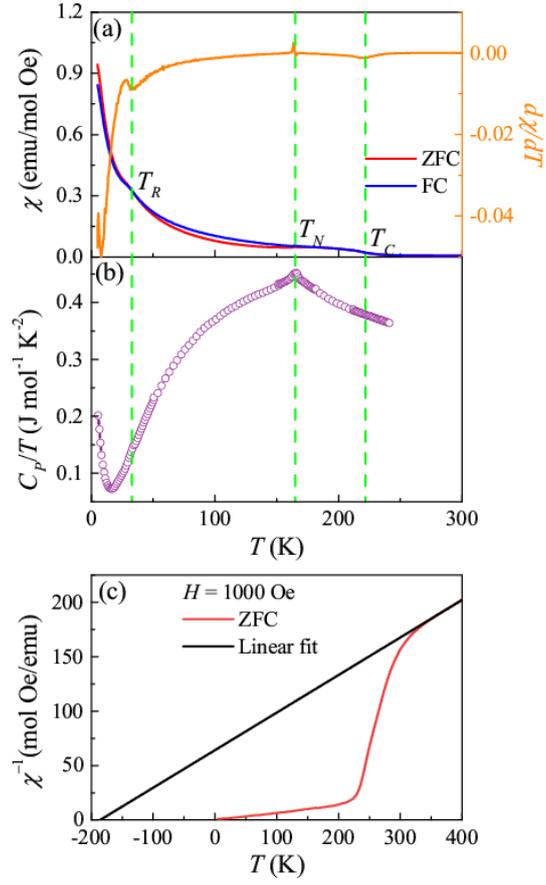

Fig. 2. (a) The *T* dependence of magnetic susceptibility ($\chi$) of Yb$_{0.42}$Sc$_{0.58}$FeO$_3$ under ZFC and FC modes measured at 1 kOe, respectively. And its temperature derivative ($d\chi/dT$) is also plotted. (b) *T*-dependent specific heat divided by *T* ($C_p/T$). (c) The Curie-Weiss fitting $\chi^{-1}$ measured under ZFC mode.

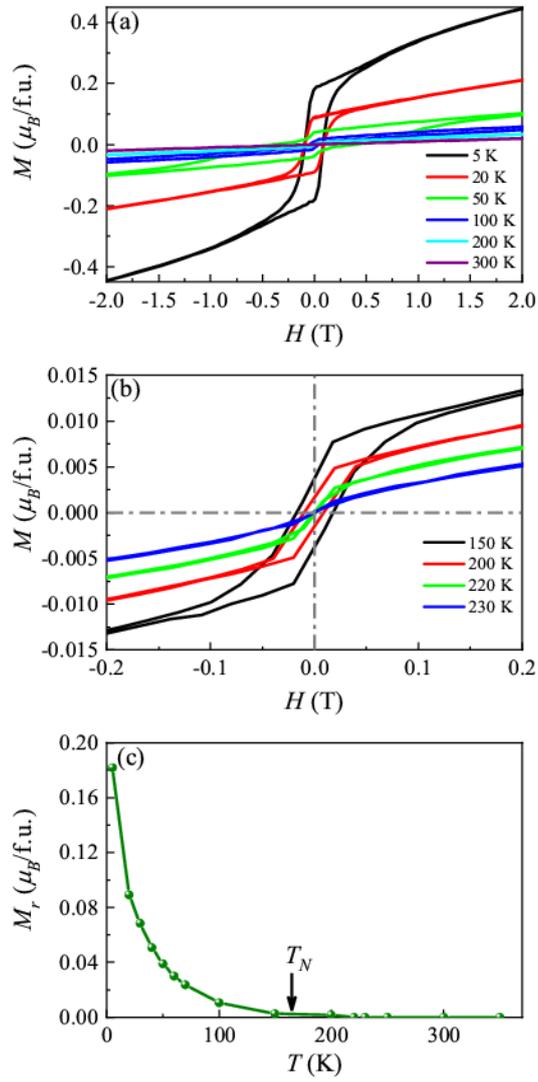

Fig. 3. (a) Magnetic field dependence of magnetization (*M*) measured at various *T*'s. (b) The *M*(*H*) curves around $T_C$ with magnetic field between -0.2 T to 0.2 T. (c) The residual magnetization ($M_r$) as a function of *T*.

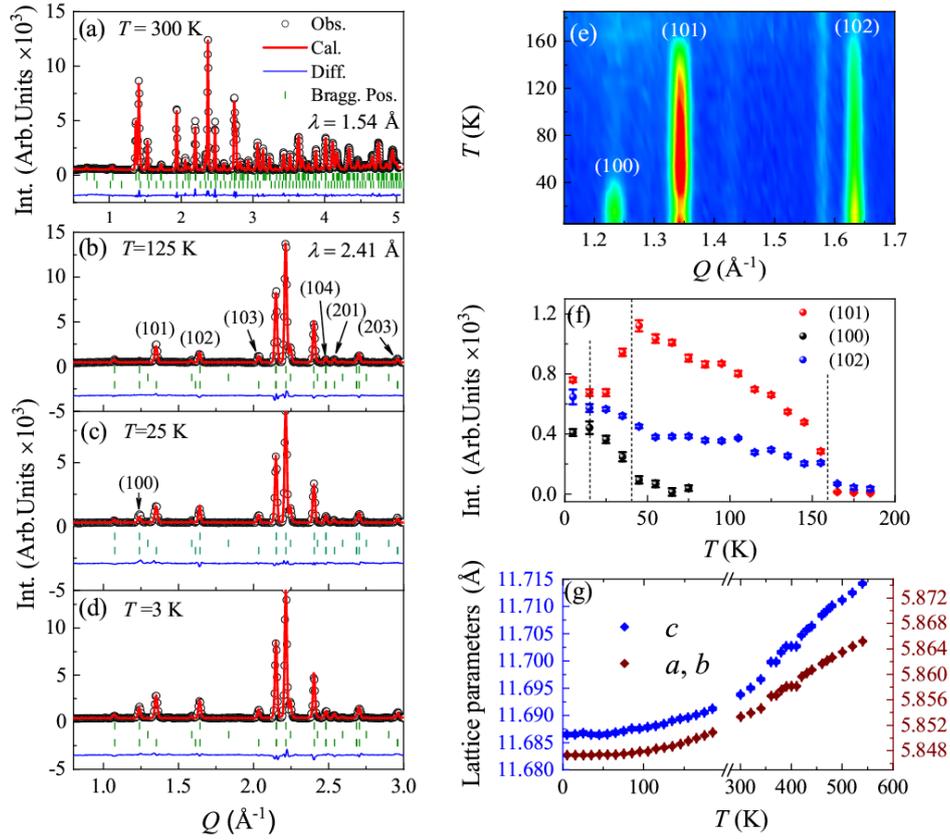

Fig. 4. The NPD patterns collected at (a) $T$ = 300 K, (b) $T$ = 125 K, (c) $T$ = 25 K, and (d) $T$ = 3 K. (e) Intensity map of the neutron-diffraction intensity about the magnetic (100), (101), and (102) reflections. (f) Integrated neutron-diffraction intensities of the magnetic (100), (101), and (102) reflections as a function of temperature. (g) Temperature dependence of refined lattice parameters $a$, $b$, and $c$ in $T$-range up to 550 K.

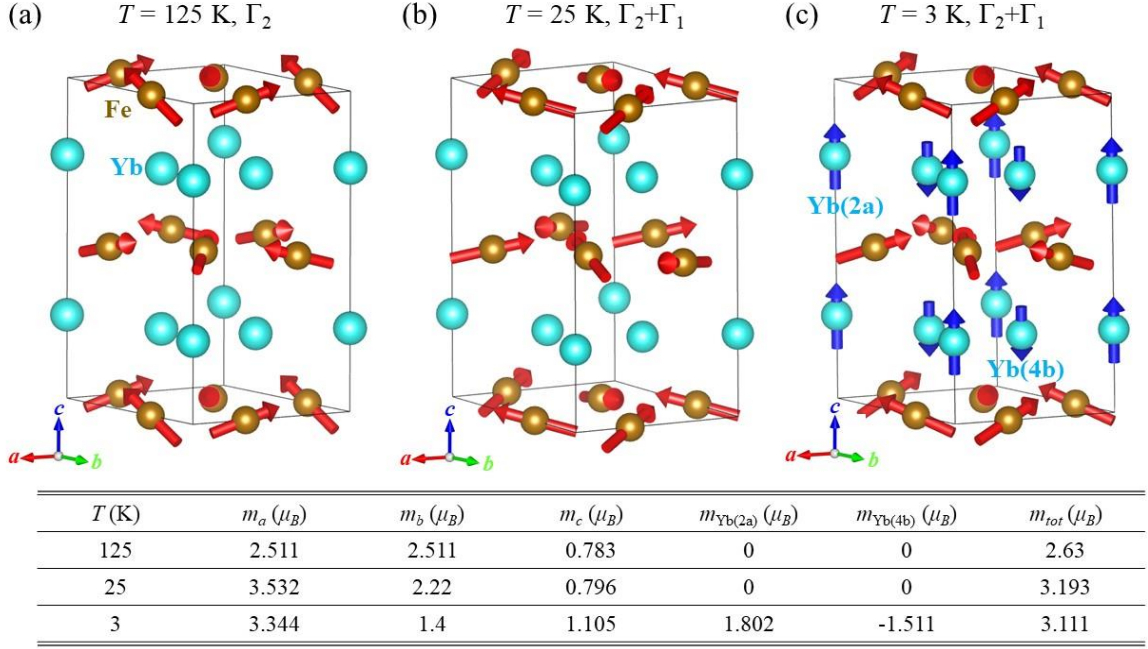

| $T$ (K) | $m_a$ ($\mu_B$) | $m_b$ ($\mu_B$) | $m_c$ ($\mu_B$) | $m_{Yb(2a)}$ ($\mu_B$) | $m_{Yb(4b)}$ ($\mu_B$) | $m_{tot}$ ($\mu_B$) |
|---|---|---|---|---|---|---|
| 125 | 2.511 | 2.511 | 0.783 | 0 | 0 | 2.63 |
| 25 | 3.532 | 2.22 | 0.796 | 0 | 0 | 3.193 |
| 3 | 3.344 | 1.4 | 1.105 | 1.802 | -1.511 | 3.111 |

Fig. 5. The refined magnetic structures at (a) 125 K, (b) 25 K and (c) 3 K, respectively. The red and blue arrows stand for the moments of $Fe^{3+}$ and $Yb^{3+}$. The magnetic moment components of $Fe^{3+}$ and $Yb^{3+}$ are listed in the table.

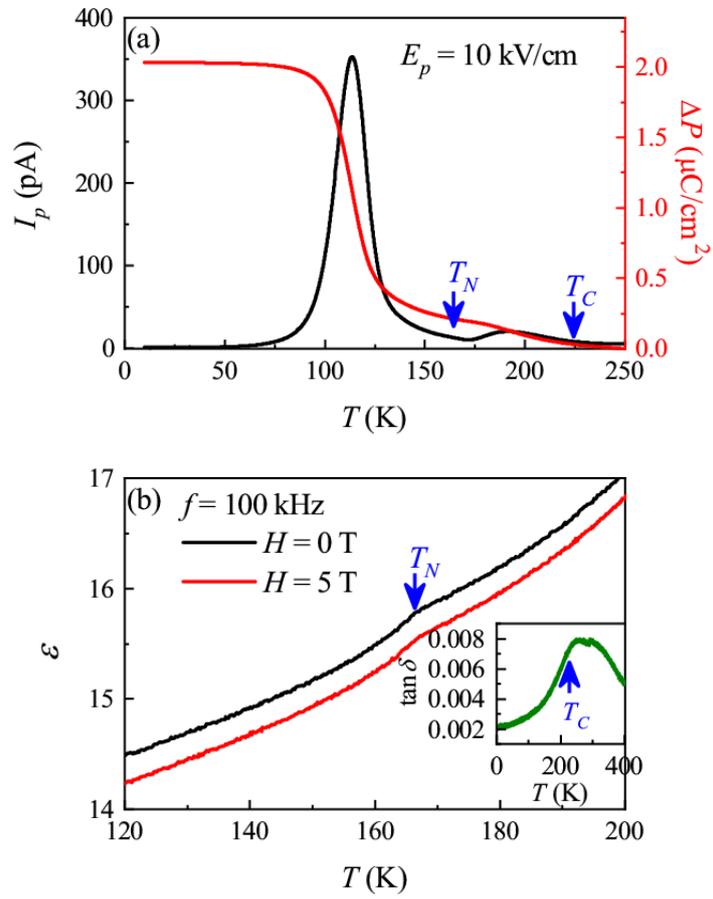

Fig. 6 (a) *T*-dependence of pyroelectric current ($I_p$) and corresponding change of polarization ($\Delta P$). (b) The *T* dependence of real part of dielectric constant. Inset: the dielectric loss.

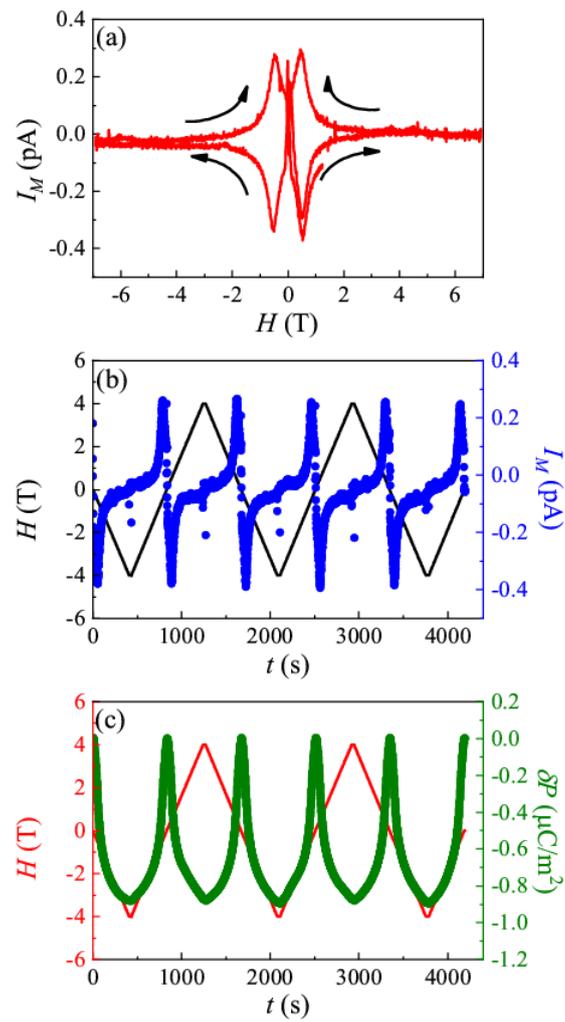

Fig.7 (a) $H$-dependence of current ($I_M$) at 2 K, measured after field cooling from 180 K. (b)(c) The ME effect repeats with $H$ oscillating between ± 4 T.

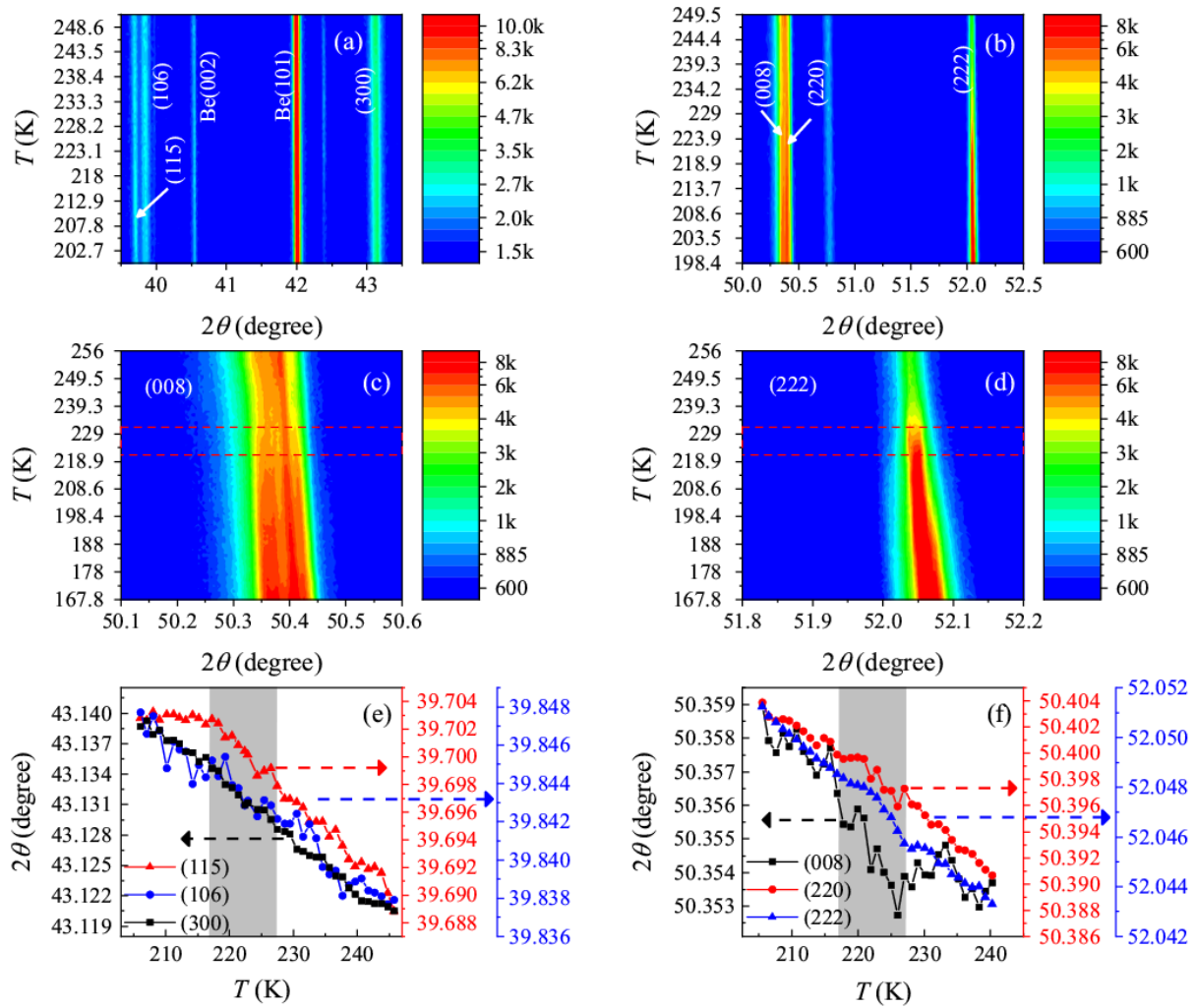

Fig. 8. *T*-dependence of XRD patterns of Yb$_{0.42}$Sc$_{0.58}$FeO$_3$ in $2\theta$ range of (a) 39.5~43.5° and (b) 50~52.5°. The angle resolution is ~0.0036°, *T* interval is 1 K, incident X-ray photon energy is 10 keV. The contour maps consist of stacking XRD patterns measured at varied *T* where the colors indicate the diffraction intensity. (c-d) The magnified view of selected regions. The *T*-dependent diffractive peak positions of (115)(106)(300)(008)(220)(222) were plotted in (e) and (f) respectively, where the color lines are guide of eyes and the green shadows indicate the *T* regions of structural distortion. For a more clear vision to the readers, the error bar of ±0.0018° was not plotted along with the data.